\documentclass[modern, a4paper, times]{aastex63}
\usepackage{amsmath}
\usepackage{graphicx}
\usepackage{natbib}

\submitjournal{ApJ}

\shorttitle{Contributions from unresolved point sources in UV maps}
\shortauthors{Strumik et al.}

\begin{document}

%\title{Separation of the heliospheric glow signal in SOHO/SWAN maps: phenomenological model, contributions from resolved and unresolved point sources}
%\title{Decomposition of SOHO/SWAN FUV maps into phenomenological heliospheric glow model and contributions from resolved and unresolved point sources}
%\title{Contributions from unresolved point sources in SOHO/SWAN UV sky maps}
\title{Inferring contributions from unresolved point sources to diffuse emissions measured in UV sky surveys: general method and SOHO/SWAN case study}

\correspondingauthor{M. Strumik}
\email{maro@cbk.waw.pl} 

\author[0000-0003-3484-2970]{M. Strumik}
\affil{Space Research Centre PAS (CBK PAN), Bartycka 18a, 00-716 Warsaw, Poland}

\author[0000-0003-3957-2359]{M. Bzowski}
\affil{Space Research Centre PAS (CBK PAN), Bartycka 18a, 00-716 Warsaw, Poland}

\author[0000-0002-6569-3800]{I. Kowalska-Leszczynska}
\affil{Space Research Centre PAS (CBK PAN), Bartycka 18a, 00-716 Warsaw, Poland}

\author[0000-0002-5204-9645]{M. A. Kubiak}
\affil{Space Research Centre PAS (CBK PAN), Bartycka 18a, 00-716 Warsaw, Poland}

\begin{abstract}
In observations of diffuse emissions like, e.g., the Lyman-$\alpha$ heliospheric glow, contributions to the observed signal from point sources (e.g., stars) are considered as a contamination. There are relatively few brightest point sources that are usually properly resolved and can be subtracted or masked. We present results of analysis of the distribution of point sources using UV sky-survey maps from the SOHO/SWAN instrument and spectrophotometry data from the IUE satellite. The estimated distribution suggests that the number of these sources increases with decreasing intensity. Below a certain threshold, these sources cannot be resolved against the diffuse signal from the backscatter glow, that results in a certain physical background from unresolved point sources.

Detection, understanding and subtraction of the point-source background has implications for proper characterization of diffuse emissions and accurate comparison with models. Stars are also often used as standard candles for in-flight calibration of satellite UV observations, thus proper understanding of signal contributions from the point sources is important for the calibration process. We present a general approach to quantify the background radiation level from unresolved point sources in UV sky-survey maps. In the proposed method, a distribution of point sources as a function of their intensity is properly integrated to compute the background signal level. These general considerations are applied to estimate the unresolved-point-sources background in the SOHO/SWAN observations that on average amounts to $28.9$ R. We discuss also the background radiation anisotropies and general questions related to modeling the point-source contributions to diffuse UV-emission observations.
\end{abstract}

%%%%%%%%%%%%%%%%%%%%%%%%%
\section{Introduction}
\label{sec:intro}

Diffuse UV radiation from various regions of the sky is an interesting problem in several space physics and astrophysical contexts, e.g., for diagnostics of astrophysical plasmas, molecular-hydrogen and dust-scattered emission. The diffuse emissions were a subject of interest of several satellite instruments, e.g., Spectroscopy of Plasma Evolution from Astrophysical Radiation (SPEAR) \citep{edelstein_etal:06} or Galaxy Evolution Explorer (GALEX) \citep{murthy:14}, providing partial or complete sky surveys. Closer to the Sun, one of the interesting problems of this kind is the Lyman-$\alpha$ (121.567 nm) heliospheric glow, which is a diffuse emission resulting from resonant scattering of solar UV radiation on hydrogen atoms around the Sun. The most complete sky-survey measurements of the characteristics of the Lyman-$\alpha$ glow were provided by the Solar Wind ANisotropy (SWAN) instrument onboard of the Solar and Heliospheric Observatory (SOHO) satellite \citep{bertaux_etal:95}.

If a diffuse emission is the main subject of interest, contributions from point sources are regarded as a contamination. Some (usually bright) point sources are seen in the sky maps as localized peaks spreading over a certain angular area depending on the angular resolution of a mapping instrument. For such point sources, peak-detection algorithms can be used to identify their position, angular extent [i.e., size in the maps corresponding to the point spread function (PSF) of a given instrument] and intensity. However, investigation of the distribution of point sources typically shows that their number increases as their intensity decreases. Below a certain threshold, faint point sources cannot be resolved against diffuse signal under investigation, even though these objects do contribute to the total signal measured. Since the contribution from such unresolved point sources cannot be distinguished from the diffuse emission, one can regard them as a background that the maps under investigation need to be corrected for. Detection, understanding and subtraction of the point-sources background has implications for proper characterization of diffuse emissions, in particular if we are interested in accurate comparison of observations with models of these emissions. Since stars are also often used as standard candles for in-flight calibration of UV observations performed by satellites, proper understanding of signal contributions from the point sources is important for the calibration process.

In this paper we present a method to derive corrections for the background emission from unresolved point sources in UV sky-survey maps. The method is based on the distribution of point sources as a function of their intensity, which can be estimated from both the sky maps of interest and from a database of observed spectra of individual astrophysical objects. The distribution is properly integrated over the range of spectral sensitivity of a mapping instrument. The integration of the distribution of point sources with appropriate limits makes it possible to estimate the background. In the implementation of our method, we used the database of the International Ultraviolet Explorer (IUE) spectra to estimate the background of the SOHO/SWAN observations of the heliospheric Lyman-$\alpha$ glow. We also discuss general questions related to point-source contributions to UV sky-surveys.

The paper is organized as follows. The proposed method of the estimation of the unresolved-point-sources background is presented in Section \ref{sec:method}. The integration and processing algorithm for the IUE spectra is described in Section \ref{sec:iue}. The proposed method is then applied in Section \ref{sec:swan} to estimate the background for the SOHO/SWAN instrument. The obtained results are discussed in Section \ref{sec:discussion}.

%%%%%%%%%%%%%%%%%%%%%%%%%
\section{Method of estimation of unresolved-point-sources background}
\label{sec:method}

Generally, the number of point sources in the range of intensities $F, F+dF$ and in a solid angle element $d\Omega$ centered at angular coordinates $l,b$ can be defined as $n(F,l,b) dF d\Omega$. The unit for the number density $n(F,l,b)$ is photon flux unit$^{-1}$ sr$^{-1}$, e.g., for the SOHO/SWAN observations the photon flux is expressed in the Rayleigh units, so the unit for $n(F,l,b)$ is R$^{-1}$ sr$^{-1}$. The Rayleigh unit definition is 1 R = $10^6$ photons s$^{-1}$ cm$^{-2}$ emitted isotropically \citep{hunten_etal:56}. Formally, such a quantity should be considered as a photon flux density, but due to a convention commonly used in the literature we will refer to it as the photon flux. If for simplicity reasons we assume that the point sources are distributed uniformly over the sky, we can use all-sky-averaged number density $n(F)$ of point sources as a function of the flux $F$ only. A discrete approximation for $n(F)$ can be found by counting the number of point sources in logarithmically-spaced photon-flux bins as follows
\begin{equation}
n(F_i)=\frac{N_i}{4 \pi \Delta_{\mathrm{F},i}},
\label{eq:nF}
\end{equation}
where $N_i$ is the number of point sources in an $i$-th bin centered at $F_i$ and $\Delta_{\mathrm{F},i}$ is the width of the bin. We can approximate the discrete estimation in Equation (\ref{eq:nF}) with a continuous function (obtained, e.g., by fitting a function in analytic form; an example containing a Schechter function fitting is presented in Figure \ref{fig:dndf}). In general, the discrete distribution in Equation (\ref{eq:nF}) can be determined in two ways. The first approach is to use a database from a programme of observations of point sources. In this paper, we use IUE observations, presented in Section \ref{sec:iue}. The second approach is to use sky-survey maps provided by a mapping instrument, identify point sources in the maps using peak-detection algorithms and then use the obtained list of point sources to compute the histogram from Equation (\ref{eq:nF}). Section \ref{subsec:crosscalibr} presents examples of using the two approaches and shows a comparison of the results.

If the effective angular area of the field of view (FOV) of the mapping instrument is $\Omega_0$, the number of the brightest point sources in the FOV with $F \ge F_0$ is given by
\begin{equation}
\hat{N}(F_0)=\Omega_0 \int_{F_0}^{F_\mathrm{cutoff}} dF ~ n(F),
\label{eq:nhat}
\end{equation}
where $F_\mathrm{cutoff}$ represents the cutoff value, which is determined by the photon flux of the brightest point source. The cutoff can be included in an analytic-form function approximating $n(F)$, then the upper limit for integration in Equation (\ref{eq:nhat}) is infinity. Using Equation (\ref{eq:nhat}) we can find a value $F_1$ for which $\hat{N}(F_1)=1$, this corresponds to exceeding a resolution limit, i.e., for $F_0<F_1$ we have more than one bright point source inside the effective angular area of the FOV of our instrument. The mean flux density of faint point sources with $F<F_0$ can be calculated as
\begin{equation}
\hat{F}(F_0)=\Omega_0 \int_{0}^{F_0} dF ~n(F) F.
\label{eq:fhat}
\end{equation}
Therefore, having beforehand determined the value $F_1$ corresponding to the resolution limit as described above, we can find $\hat{F}(F_1)$ which provides an estimation of the average flux from faint unresolved point sources.

In the discussion presented above we refer to the effective angular area $\Omega_0$ of the mapping instrument, which requires further explanation. In analogy to classical optical microscopy, where the Rayleigh criterion is used to estimate the resolution limit, we assume that two point sources are properly resolved if they can be distinguished as individual objects in a light curve traversing their positions. If two point sources with the same photon flux are closely located, the light curve has two maxima corresponding to the positions of the point sources and a local minimum between them. The Rayleigh criterion implies that the point sources are properly resolved if the contrast between the local minimum and the neighboring local maxima is at least 26\%. We adopt this threshold value for our definition of the effective angular area, which in the case of circular FOV will be equal to the spherical cap area $\Omega_0=2\pi(1-\cos\rho_\mathrm{res})$, where the angular radius $\rho_\mathrm{res}$ corresponds to the marginal case of the maxima-minimum contrast of 26\%. The value of $\rho_\mathrm{res}$ will depend on the shape of the PSF of the mapping instrument, which describes the detector response for a point source shifted by an angle off the collimator axis. An example of the estimation of $\rho_\mathrm{res}$ for the SOHO/SWAN instrument is presented in Section \ref{subsec:psf}.

%%%%%%%%%%%%%%%%%%%%%%%%%
\section{IUE spectra integration and processing}
\label{sec:iue}
The IUE satellite executed a programme of selective spectrophotometric observations of UV point sources in the range of wavelengths between 115 and 320 nm. The measurements were obtained from 1978 to 1996. In this paper we analyze data from the Mikulski Archive for Space Telescopes IUE archive \citep{mast:iue} that contains absolutely-calibrated fluxes, data quality flags, and measurement uncertainties as a function of the wavelength. Some data files are provided in high resolution (MXHI) and the other in low resolution (MXLO). 

The analyzed MXHI and MXLO spectra are provided in the IUE database as arrays, where absolutely-calibrated fluxes $E_j$ (given in erg cm$^{-2}$ s$^{-1}$ \AA$^{-1}$ units), data quality flags $q_i$, and measurement uncertainties $\sigma_j$ are given as a function of discrete-grid wavelength values $\lambda_j$ for the range of wavelengths 115-198 nm, where $j= 1, \ldots, N_\lambda$ and $N_\lambda$ is the size of the wavelength array. For the MXLO spectra the wavelength grid with $N_\mathrm{MXLO}=640$ points is regular (in the sense of equal sampling step), whereas MXHI spectra are provided on a set of regular grids, where each set contains $N_\mathrm{MXHI}=768$ points, but the step for wavelength sampling in each set may be generally different. The MXHI spectra show also strong signatures of contaminations by uncorrelated noise.

Our postprocessing of the IUE spectra involves smoothing them by the second-order spline approximation and then resampling by linear interpolation on a regular grid with the wavelength increment of $0.01$ nm. These steps are expected to tackle the problem of noise and to make the wavelength grid regular for the MXHI files. We also apply another quality-checking postprocessing step, where we verify coordinates and target name in the MXLO and MXHI files against the SIMBAD astronomical database. If the right ascension or declination differ by more than 0.05 deg from the database values, we skip a given spectrum file.

For a UV mapping instrument, a spectral sensitivity $\eta(\lambda)$ curve is normally known from ground calibration measurements. An interpolation procedure can be used to obtain the spectral sensitivity values $\eta_j$ corresponding to the wavelength values $\lambda_j$. Then the expected flux for an $i$-th point source can be computed as
\begin{equation}
F_i=A \sum_{j=1}^{N_\lambda} \eta_j E_{j,i} \Delta_\lambda.
\label{eq:Fint}
\end{equation}
The factor $A$ normally represents the effective area of a detector. However, we can set $A=1$ and try to find a cross-calibration factor between the IUE observations and photon flux measured by a given instrument, as discussed below for the SOHO/SWAN observations. Equation (\ref{eq:Fint}) is applied to all MXLO and MXHI files in the IUE database, which makes it possible to obtain the photon fluxes.

For a number of point sources, multiple IUE spectra measurements are available (both MXLO and MXHI files) and the point sources may in general emit time-variable flux. To tackle this problem, from multiple spectra we select the average flux for consistency with the SOHO/SWAN point-source identification procedure described in Section \ref{sec:swan}. If for a given target both MXLO and MXHI spectra are available, the fluxes from these two types of spectra are averaged.

%%%%%%%%%%%%%%%%%%%%%%%%%
\section{Analysis of the SOHO/SWAN observations}
\label{sec:swan}

The SWAN instrument onboard the SOHO satellite is an all-sky mapper devoted to observations of the Lyman-$\alpha$ backscatter heliospheric glow \citep{bertaux_etal:95}. The glow is generated as a result of scattering of the solar Lyman-$\alpha$ photons on the local interstellar hydrogen atoms. As interstellar hydrogen atoms penetrate the heliosphere, a cavity in their density distribution forms around the Sun due to charge exchange with the solar-wind protons and photoionization processes. Due to variations in the hydrogen density, ionization rate, radiation pressure, as well as variations in the solar Lyman-$\alpha$ emissions, the Lyman-$\alpha$ glow is modulated in the sky and varies in time.

\begin{figure}[!htbp]
\begin{center}
\includegraphics[scale=0.5]{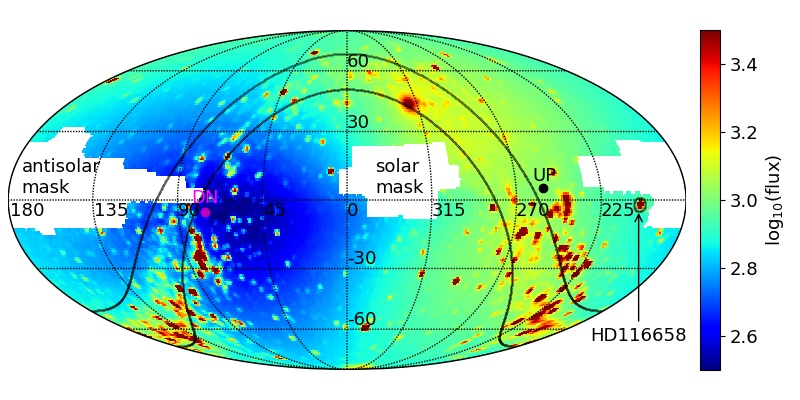}
\caption{An example of the SOHO/SWAN map obtained on 18 February 1997. The map is shown in the ecliptic coordinates using the Mollweide projection. Dipole-like pattern of the heliospheric glow has its maximum in the proximity of the upwind direction (black circle marked as UP) and the minimum near the downwind direction (magenta circle marked as DN). On the slowly-varying heliospheric-glow background, point sources can be identified as local peaks of the photon flux. An example point source HD116658 is indicated by the arrow and surrounded by the 3-deg circle. A galactic band is shown between two black lines located $\pm 10$ deg from the galactic equator. White regions in the map show the location of the solar and antisolar masks.}
\label{fig:swan_map_example}
\end{center}
\end{figure}

The SOHO/SWAN maps are provided as daily scans of the sky with 1-deg angular resolution in the ecliptic coordinates. The modulation of the observed photon flux density is provided in the Rayleigh units. Although the instrument is devoted to the studies of the Lyman-$\alpha$ glow, its spectral sensitivity range extends from 115 to 185 nm. For this reason the instrument is quite sensitive to non-Lyman-$\alpha$ emissions from stars. Fig. \ref{fig:swan_map_example} shows an example of the SOHO/SWAN map obtained on 18 February 1997, based on a FITS file from the SOHO/SWAN database \citep{soho:swan}.
To make both point sources and the heliospheric glow more visible, the map shows the logarithm of the photon flux density. In the map, we show the local-interstellar-medium upwind and downwind directions, the galactic band, an example point source, and white masked regions. The upwind direction determined by \citet{bzowski_etal:15a} is used here, where the ecliptic longitude is 255.8 deg, and the ecliptic latitude 5.16 deg. The white regions were masked by the instrument team, because valid data is not available close to the solar and anti-solar direction. In these masked regions, measurements during a given day were not possible due to solar stray light or satellite reflections of the solar light.

For the heliospheric glow studies, the photon flux from stars and the Galaxy is considered as a contamination that impedes a direct comparison of the glow with models. In this section the general method presented in Section \ref{sec:method} is applied to determine the unresolved-point-sources background in the SOHO/SWAN maps.

\subsection{Extracting point-sources map from SOHO/SWAN observations}
As presented in Fig. \ref{fig:swan_map_example}, maps provided by the SOHO/SWAN instrument contain masked regions and the modulated heliospheric glow, which makes immediate analysis of the contribution from point sources difficult. Therefore, for further analysis it is advantageous to generate time-averaged SOHO/SWAN maps with the heliospheric-glow background subtracted. Subtracting the background is not a trivial task because the glow features a slowly-varying modulation pattern in the sky. The following procedure was applied to get glow-detrended maps.

\begin{figure}[!htbp]
\begin{center}
\includegraphics[scale=0.5]{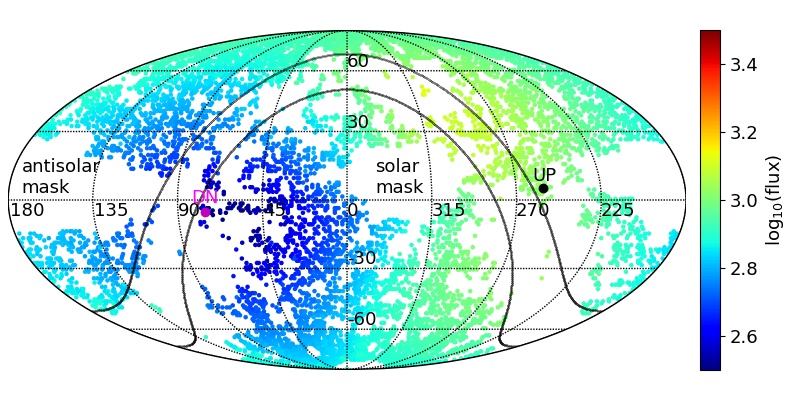}
\caption{An example of the SOHO/SWAN map, where an additional signal-variability-masking procedure has been applied for approximately $15000$ randomly distributed pixels from Figure \ref{fig:swan_map_example}. The map shows approximately $8600$ points with low-$\sigma$ neighborhood, that are then used to fit a heliospheric-glow approximator.}
\label{fig:swan_map_example_mask}
\end{center}
\end{figure}

For a given daily map we identify pixels with highly-variable neighborhood (the standard deviation of the signal $\sigma>30$ R in the 3-deg circle around a given pixel) and mask them out as regions, where resolved point sources are supposed to give substantial contribution. This criterion is checked for a limited number (approximately 15000) of pixels uniformly distributed over the map to reduce the computational cost of fitting an approximator described further. The result of masking is shown in Figure \ref{fig:swan_map_example_mask}, where approximately 8600 pixels with low-$\sigma$ neighborhood are shown in color.
By comparison of Figures \ref{fig:swan_map_example} and \ref{fig:swan_map_example_mask} we can identify the additional signal-variability-related mask in addition to the solar and antisolar masks. A significant part of the sky adjacent to the galactic equator has been masked out as a result of using the criterion described above, because this region contains many point sources. The remaining points with low-$\sigma$ neighborhood are then used to fit a function that slowly varies over the sky – this is assumed to be a phenomenological heliospheric glow approximator. For this purpose the support vector regression approximating function svm.SVR(kernel='rbf',C=50000.0,gamma=1.5) was adopted from a machine-learning software package scikit-learn \citep{pedregosa_etal:11}. The parameters C and gamma determine the flexibility and the characteristic scale of variability of the approximator in the sky. The parameter values given above were found as providing good separation of small- and medium-scale features from large-scale features in the SOHO/SWAN maps. As the small-scale features we refer to the resolved point sources, medium-scale -- the band of increased photon flux in the proximity of the galactic equator, and the large-scale feature is the heliospheric glow.

The approximation essentially consists in fitting an ensemble of local radial-basis-function approximators that is conceptually similar to commonly-used spline-function detrending. More details can be found in the scikit-learn package documentation of the approximating function \citep{scikit-learn:svr}. Using this procedure we essentially find a baseline or a lower envelope of the signal that is expected to represent the heliospheric glow. Everything above the baseline is interpreted as a contamination. Obviously, the baseline should be understood as a surface because we operate on two-dimensional maps.

\begin{figure}[!htbp]
\begin{center}
\includegraphics[scale=0.55]{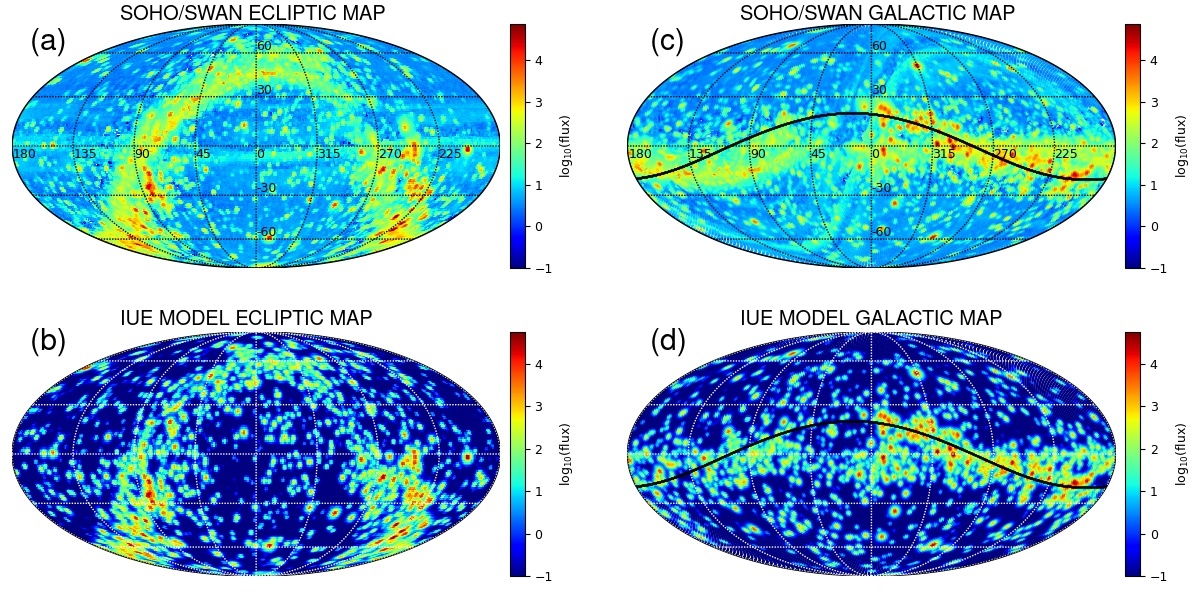}
\caption{Point sources in full-sky SOHO/SWAN maps as (a,c) observed and (b,d) modeled using the IUE stars observations. All panels show the logarithm of the flux in Rayleigh units. The Mollweide projection and two frames are used for the visualization: (a,b) the ecliptic and (c,d) galactic coordinates. In panels (a,c), the point-source features were boosted by appropriate postprocessing of the SOHO/SWAN maps, consisting in subtracting the heliospheric glow and averaging the maps over the year 1997. In panels (b,d), spectra for 5615 observed point sources from the IUE catalogue were integrated over a range of the spectral sensitivity of the SOHO/SWAN instrument, which makes it possible to generate the expected map of the signal from stars using an estimated PSF for the SOHO/SWAN instrument. Black line in panels (c,d) shows a great circle crossing the galactic equator at $\sim\!\!20$ deg with its ascending node at $\sim\!\!285$ deg, indicating the position of the Gould Belt.}
\label{fig:swan_map}
\end{center}
\end{figure}

The heliospheric glow approximators are then subtracted from the daily maps. There are regions in the maps masked by the instrument team, but since the satellite orbits around the First Lagrangian Point (L1), it moves around the Sun with the Earth and the masked regions move in the sky. Therefore, to obtain a full-sky map we computed an average map over the year 1997, where the masked pixels were excluded from the averaging. The postprocessing procedure presented above results in a map shown in Figure \ref{fig:swan_map}(a).

A significantly larger number of point sources is seen as localized small-scale spots in Figure \ref{fig:swan_map}(a) as compared with Figure \ref{fig:swan_map_example}. The map from Figure \ref{fig:swan_map}(a) transformed to the galactic coordinates is shown in Figure \ref{fig:swan_map}(c). The point sources are seen as localized peaks, most of them located in the proximity of the galactic plane and seen as the bright band. The faint narrow stripe around the ecliptic equator appears as a trail of slightly brighter spots on the edge of the masked regions, as the masks are moving in the sky during the year.

Another feature that can be identified in the maps of Figure \ref{fig:swan_map}(a) and (c) is a statistical scatter of the heliospheric glow seen as the light-blue all-sky background. The statistical scatter appears as a result of the processing of the maps, where first the smooth slowly varying component is subtracted, then the resulting sequence of maps is averaged over one year. The relative accuracy of the SOHO/SWAN instrument for the interplanetary signal is $\sim\!\!1\%$ as described in Table I by \citet{bertaux_etal:95}. Since typical intensity of the heliospheric glow measured by the SOHO/SWAN instrument is 400-1000 R, the expected statistical scatter level is several R, which is indeed observed in Figure \ref{fig:swan_map}(a). The statistical scatter is investigated in a more detail further in the paper.

Figures \ref{fig:swan_map}(b) and (d) contain maps based on the IUE observations, that can be considered as an IUE-based model of the point sources seen by the SOHO/SWAN instrument. Detailed description of a procedure used for generation of the model maps is included below.

\subsection{Estimation of the PSF and the effective resolution}
\label{subsec:psf}
\begin{figure}[!htbp]
\begin{center}
\includegraphics[scale=0.55]{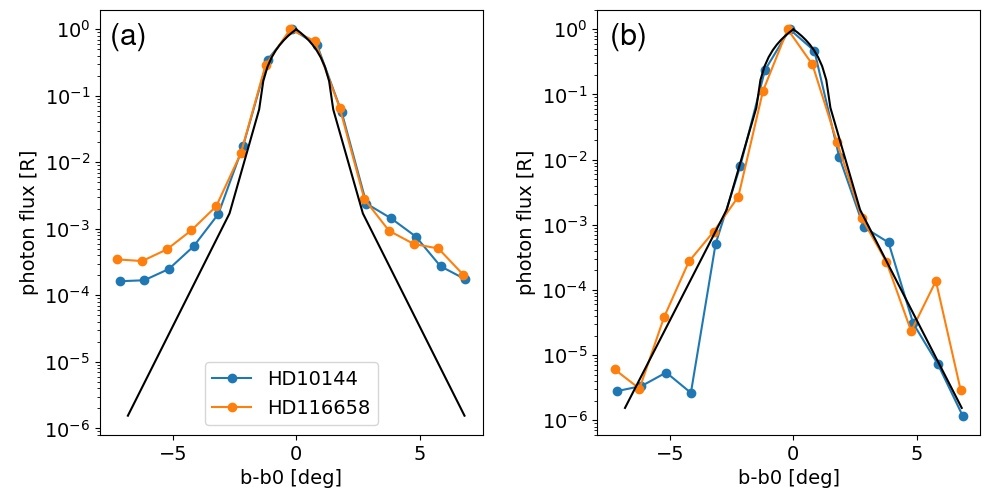}
\caption{Estimation of the PSF based on observations of two stars that are expected to be well separated from other point sources. Latitudinal dependence of the signal is shown (blue and orange), where the light curves are normalized to their respective maxima. The estimated PSF (black line) was chosen to fit reasonably the core of the point-source observations in (a) yearly-averaged map and wings seen in (b) yearly-minimum measurements.}
\label{fig:psf}
\end{center}
\end{figure}

By subtracting the heliospheric glow approximator from raw SOHO/SWAN maps we can better extract point sources. Averaging over one year gives us more stable and robust estimations for extra-heliospheric point sources as the effects of transient events (e.g., passage of comets) and statistical scatter on the maps are minimized. Since stars can be considered as point sources, analysis of images of stars in the postprocessed SOHO/SWAN maps allows us to determine the instrument PSF, as illustrated in Figure \ref{fig:psf}.
Due to the presence of the statistical scatter in the yearly-averaged map presented in Figures \ref{fig:swan_map}(a) and (c) only the core of the PSF can be reliably determined, because the statistical scatter elevates the wings of the signal around stars, as illustrated in Figure \ref{fig:psf}(a). This effect can be mitigated to some extent by using yearly-minimum observations, as shown in Figure \ref{fig:psf}(b). Therefore, the estimated PSF function was chosen as a reasonable trade-off that fits the core of the point-source observations in yearly-averaged map and the wings seen in yearly-minimum measurements.

Naturally, the SOHO/SWAN detector boresight is subject to pointing uncertainties that lead to day-to-day variations of the signal from point sources as mapped onto individual pixels of the SOHO/SWAN maps. This is the reason why the averaging procedure described above gives slightly larger point-source spread in yearly-averaged case in Figure \ref{fig:psf}(a) as compared with the yearly-minimum case (b) for the core of the PSF. Based on our experience with the analysis of daily SOHO/SWAN maps, the pointing-uncertainty effects become more pronounced at times in certain parts of the sky, which leads to slight systematic increase of the angular extent of point sources in yearly-averaged maps. In our analysis presented below we bear these effects in mind when presenting the discussion of various results related to signal contributions from point sources. In particular, comparisons of modeled distribution of point sources and SOHO/SWAN observations presented below suggest that the pointing-uncertainty effects do not influence conclusions presented in our paper.

The estimated PSF in the proximity of its maximum varies like $PSF(\rho) \approx 1.0-|\rho/1.6|$, which leads to the resolution limit $\rho_\mathrm{res} \approx 2$ deg if the definition specified in Section \ref{sec:method} is used. The maps obtained from the SOHO/SWAN measurements are binned with 1-deg angular resolution. Therefore, we can assume that to resolve two point sources we need them to be separated by at least 2 deg in the latitudinal or longitudinal direction (i.e., we need at least one pixel between two neighboring pixels to confirm the presence of a local minimum between two local maxima). The required separation increases, if resolving along pixel diagonal is taken into account. For further considerations we adopt the latter (diagonal) value as the angular separation defining the resolution threshold $\rho_\mathrm{res}\approx 2.83$ deg, which corresponds to the worst-case resolving scenario. Assuming the circular shape of the FOV we get the effective angular area $\Omega_0=2\pi(1-\cos\rho_\mathrm{res}) \approx 7.66 \times 10^{-3}$ sr for the analysis of the resolution-related issues.

\subsection{IUE-SOHO/SWAN cross-calibration factor and modeling distribution of point sources in SOHO/SWAN observations}
\label{subsec:crosscalibr}
Using the spectral sensitivity curve for the SOHO/SWAN instrument [Figure 13.4 in \citet{quemerais_bertaux:02}] as an estimation of $\eta_j$ and absolutely-calibrated fluxes $E_j$ from the IUE spectra described in the previous section, from Equation (\ref{eq:Fint}) we computed the expected photon fluxes $F_i$ for the IUE stars. Since the SOHO/SWAN spectral sensitivity curve is provided simply as the counting rate during a ground calibration experiment and the radiation source characteristics are not known in detail, the flux $F_i$ we obtain can be regarded as given in arbitrary units and we need a calibration factor to transform the SOHO/SWAN-adjusted fluxes $F_i$ of the IUE stars from the arbitrary units to the Rayleigh units that are used in the SOHO/SWAN maps.

Due to rather low resolution of the SOHO/SWAN observations, it is a quite common situation that for an observed star, the measured photon flux will be increased due to stray light from other nearby stars within the effective FOV. To mitigate this problem in our cross-calibration procedure, we computed the expected map of point sources in the ecliptic coordinates shown in Figure \ref{fig:swan_map}(b) and in the galactic coordinates as presented in \ref{fig:swan_map}(d). The maps were computed using 5615 stars from the IUE observation catalogue and the PSF function estimated in the previous section, to obtain possibly realistic photon flux and size of the point sources in the map.

\begin{figure}[!htbp]
\begin{center}
\includegraphics[scale=0.8]{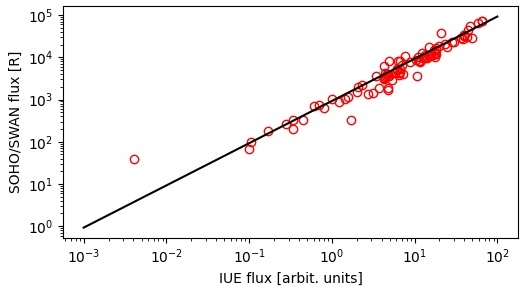}
\caption{Estimation of the IUE-SOHO/SWAN cross-calibration factor. Red circles represent values of the flux for each star derived from two approaches: integration of the IUE spectra followed by using the PSF (horizontal axis) and direct observation by the SOHO/SWAN instrument (vertical axis). Black line shows the best fit from linear regression.}
\label{fig:calibr}
\end{center}
\end{figure}

One should note that the maps in Figures \ref{fig:swan_map}(b) and (d) are already presented in Rayleigh units, as obtained using a IUE-SOHO/SWAN cross-calibration factor described below. For the cross-calibration purposes we use 91 stars listed in Table 7.4 by \citet{snow_etal:13}. Results of the calibration are shown in Figure \ref{fig:calibr}.
The cross-calibration factor was found equal to $\sim\!\!$857.59 as the best fit from linear regression in the logarithmic scale. The factor should be understood as a coefficient relating the IUE flux in the arbitrary units, obtained from integration of the convolution of the IUE spectra and the SOHO/SWAN spectral sensitivity using Equation (\ref{eq:Fint}), with the SOHO/SWAN flux in the Rayleigh units.

\begin{figure}[!htbp]
\begin{center}
\includegraphics[scale=0.8]{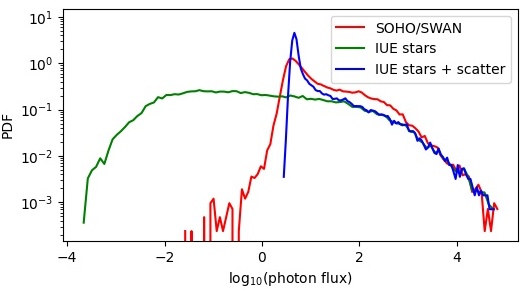}
\caption{Histogram of the maps presented in Figures \ref{fig:swan_map}(a) (red line) and \ref{fig:swan_map}(b) (green line). Blue line shows the histogram for Figure \ref{fig:swan_map}(b) with Poisson-distributed statistical scatter of a synthetic glow signal added.}
\label{fig:map_histogram}
\end{center}
\end{figure}

The maps shown in Figures \ref{fig:swan_map}(b) and (d) can be considered as a model of the intensity of known point sources seen by the SOHO/SWAN instrument based on the IUE spectra measurements. The total observed intensity of the sky is shown in Figures \ref{fig:swan_map}(a) and (c), correspondingly. One can see a general similarity between the corresponding maps shown in Figure \ref{fig:swan_map}, suggesting that the contribution from point sources is reasonably modeled. One of the differences is the presence of diffuse contribution in the proximity of the galactic plane in Figures \ref{fig:swan_map}(a) and (c), seen as light green color between stars in the galactic band, which is not so pronounced in Figures \ref{fig:swan_map}(b) and (d). A possible source of the diffuse contribution can be a real diffuse signal from the galactic plane and/or signal from point sources that were not included in the IUE observation programme, since the programme was not a regular survey of the sky. Another difference is the presence of the all-sky background from the statistical scatter of the heliospheric glow in Figures \ref{fig:swan_map}(a) and (c) (light blue color), which is not seen in Figures \ref{fig:swan_map}(b) and (d) as it is not included in the model used to produce the map. Figure \ref{fig:map_histogram} shows histograms of the maps from Figure \ref{fig:swan_map},
where the influence of the statistical scatter of the heliospheric glow is presented. The statistical-scatter effect was included here by adding to the map in Figure \ref{fig:swan_map}(b) a synthetic component corresponding to the expected Poisson-distributed noise for the counting rate for the heliospheric glow set to a constant flux of 400 R. The difference between the blue and green lines in Figure \ref{fig:map_histogram} illustrates the effects of the statistical scatter, making the IUE-stars-with-statistical-scatter histogram (blue line) similar to the SOHO/SWAN histogram (red line). Here we used a constant flux of 400 R for the heliospheric glow which is a strongly simplifying assumption that leads to narrower peak in the blue line as compared to the red line, where full variability of the heliospheric glow signal is naturally included. However, since detailed modeling of the heliospheric glow is beyond the scope of the present paper, we present only the simplified model illustrating the effects of the statistical scatter qualitatively.

\begin{figure}[!htbp]
\begin{center}
\includegraphics[scale=0.6]{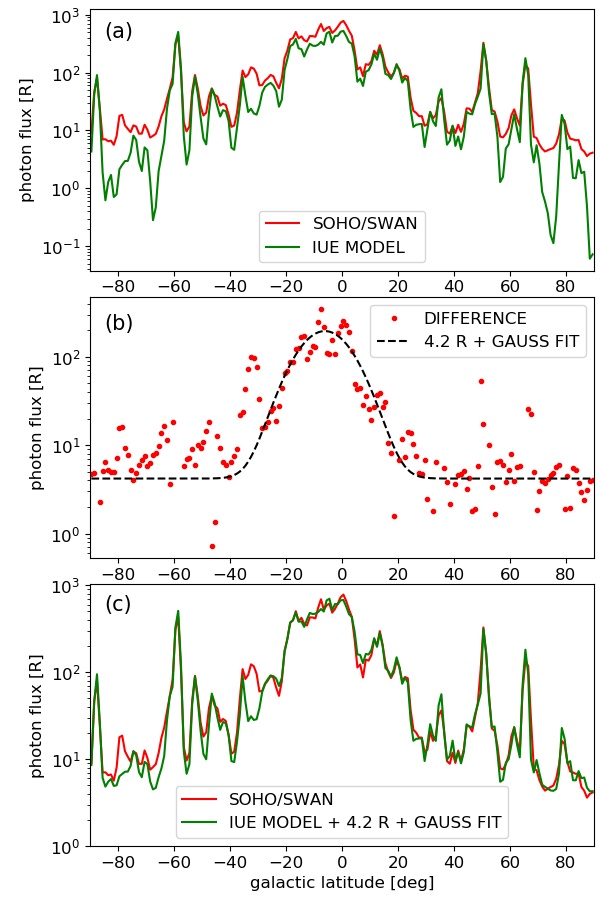}
\caption{Galactic-longitude-averaged distribution of the signal from point sources. Panel (a) shows comparison of the SOHO/SWAN observations with raw IUE-based model. In panel (b) the difference between the curves shown in panel (a) is fitted with the superposition of a constant background and a Gaussian function. The corrected model shown in panel (c) shows much better agreement with the SOHO/SWAN observations as compared to the raw model presented in panel (a).}
\label{fig:swan_map_long_avg}
\end{center}
\end{figure}

Figures \ref{fig:swan_map}(c) and (d) suggest that the maximum of the distribution of the signal from point sources is located near the galactic equator. To investigate this pattern in a more detail, in Figure \ref{fig:swan_map_long_avg}(a) we present the 
distribution of the signal averaged over the galactic longitude. The plot shows that the raw IUE-based model predicts rather systematically smaller values of the signal as compared to the SOHO/SWAN observations, which is related to the selective character of the IUE observations. Although the maximum of the longitudinal average of the signal is located close to the equator, some asymmetries can be identified by looking at the difference between the SOHO/SWAN observations and the IUE-model as presented in Figure \ref{fig:swan_map_long_avg}(b). The difference can be approximated by a constant background of $\sim\!\!4.2$ R and a Gaussian-distributed signal component in the proximity of the galactic equator. The constant background corresponds to the statistical scatter for the heliospheric glow, corresponding to the peak for the red line in Figure \ref{fig:map_histogram}. The Gaussian-distributed component of the angular width $\sim\!\!9.21$ deg and the amplitude $\sim\!\!191.1$ R is centered at $\sim\!\!-6.11$ deg. Figure \ref{fig:swan_map_long_avg}(c) shows a comparison of the SOHO/SWAN observations and the IUE-based model with the components shown in Figure \ref{fig:swan_map_long_avg}(b) with black dashed line included. One can see a good correspondence between the SOHO/SWAN observations of point sources and the corrected IUE-based model.

\subsection{All-sky-averaged number density of point sources and average flux from unresolved point sources}
\label{sec:avg_flux}
\begin{figure}[!htbp]
\begin{center}
\includegraphics[scale=0.8]{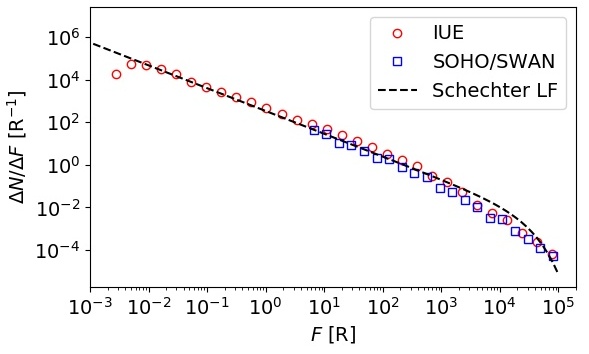}
\caption{All-sky-averaged number density of point sources estimated from IUE observations (red circles) and SOHO/SWAN maps (blue squares). Black dashed line shows the best fit of the IUE data with the Schechter luminosity function.}
\label{fig:dndf}
\end{center}
\end{figure}

Using the SOHO/SWAN-adjusted photon fluxes of the IUE stars we computed the histogram presented in Figure \ref{fig:dndf} representing the number density of point sources as discussed for Equation (\ref{eq:nF}). The distribution was computed in two ways. In the first approach, we used 5615 stars from the IUE observations catalogue, that allowed us to obtain almost eight orders of magnitude of variability of the photon flux (see red circles). In the second approach, we used the map shown in Figure \ref{fig:swan_map}(a) to identify point sources using peak-finding algorithms, that gives more than four orders of magnitude for the photon flux (see blue squares in Figure \ref{fig:dndf}). We found a good agreement between the results provided by the two approaches.

One should note that the histogram presented in Figure \ref{fig:dndf} conceptually corresponds to a luminosity function used as a statistical tool in studies of the evolution of galaxies (see, e.g., \citet{johnston:11}). A commonly used analytic-form approximation for the luminosity function is the Schechter function
\begin{equation}
\phi(F)=\frac{\psi_\ast}{F_\ast} \left( \frac{F}{F_\ast} \right) ^\alpha \exp \left( -\frac{F}{F_\ast} \right),
\end{equation}
where $F_\ast$ determines a break in the luminosity function, i.e., the photon flux at which the power-law dependence alters into the exponential cut-off. An attempt of fitting this approximation to the IUE-based data in Figure \ref{fig:dndf} (red circles) gives the following parameter values $\psi_\ast \approx 167.35$, $F_\ast \approx 18775.5$, $\alpha\approx-1.07$.
One can see some discrepancies on the left, for the smallest values of the photon flux, that we interpret as related to the selection effects in the IUE observations. Although the Schechter-function fit looks reasonably as an approximation of the IUE distribution, generally some systematic deviations are seen. Namely, for $0.2<F<500$ R the red circles are slightly above the Schechter approximation and for $2000<F<40000$ R they are slightly below. One must remember that the presented logarithmic-scale plot extends over 10 orders of magnitude in the vertical axis, thus even a small discrepancy between the fitted function and the IUE data may give significant effects, especially for integrals, as discussed below.

By fitting the observed distribution in Figure \ref{fig:dndf} with the Schechter function we can make some analytical estimations of the contributions from the unresolved sources to the SOHO/SWAN maps. According to Equation (\ref{eq:nhat}) we get
\begin{equation}
\hat{N}(F_0)=\frac{\Omega_0}{4\pi} \int_{F_0}^{\infty} dF ~ \phi(F) = \frac{\Omega_0 \psi_\ast}{4\pi} \Gamma \left( 1+\alpha,\frac{F_0}{F_\ast} \right),
\label{eq:nhat_schechter}
\end{equation}
where $\Gamma(a,x)$ is the upper incomplete gamma function
\begin{equation}
\Gamma(a,x)=\int_x^\infty t^{a-1} e^{-t} dt.
\end{equation}
Equation (\ref{eq:nhat_schechter}) provides a Schechter-function-based estimation of the number of bright point sources with $F \ge F_0$ in the FOV of angular area $\Omega_0$. Similarly, using Equation (\ref{eq:fhat}) we can compute the average photon flux from faint point sources with $F \le F_0$
\begin{equation}
\hat{F}(F_0)=\frac{\Omega_0}{4\pi} \int_{0}^{F_0} dF ~ F ~ \phi(F) = \frac{\Omega_0 \psi_\ast F_\ast}{4\pi} \left[ \Gamma \left( 2+\alpha,0 \right) - \Gamma \left( 2+\alpha,\frac{F_0}{F_\ast} \right) \right].
\label{eq:fhat_schechter}
\end{equation}

\begin{figure}[!htbp]
\begin{center}
\includegraphics[scale=0.8]{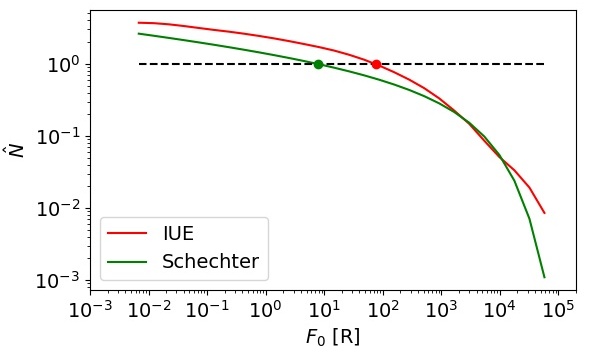}
\caption{The number of bright point sources in the FOV of the SOHO/SWAN instrument obtained by integration of the Schechter function (green) and direct numerical integration of the IUE data (red). Dots show estimated values of the photon flux $F_1$ corresponding to the resolution limit.}
\label{fig:nhat}
\end{center}
\end{figure}

Figure \ref{fig:nhat} shows a comparison of the Schechter-function-based estimation of $\hat{N}(F_0)$ with results of direct numerical 
integration of the IUE data, using the FOV area $\Omega_0$ estimated in Section \ref{subsec:psf}. This plot allows us to find the value of $F_1$ corresponding to the resolution limit, where $\hat{N}(F_1)=1$ (for $F_0 < F_1$ we have more than one bright point source in the FOV of the SOHO/SWAN instrument). We can see that previously discussed discrepancies between the IUE data (red line) and the Schechter-function fit (green line) are strengthened by the integration process and lead to significant deviations of the estimated value $F_1$, which for the Schechter-function integral is $F_1 \approx 7.8$ R (green dot) and for the direct numerical integration of the IUE data we get $F_1 \approx 76.5$ R (red dot).

\begin{figure}[!htbp]
\begin{center}
\includegraphics[scale=0.8]{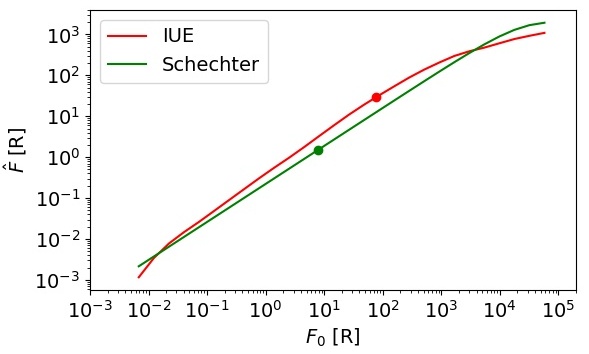}
\caption{Average flux from faint point sources in the FOV of the SOHO/SWAN instrument obtained by integration of the Schechter function (green) and direct numerical integration of the IUE data (red). Dots show estimated values of the photon flux from faint unresolved point sources.}
\label{fig:fhat}
\end{center}
\end{figure}

In Figure \ref{fig:fhat} we show a plot of the average flux $\hat{F}(F_0)$ from faint point sources for the Schechter-function 
approximation (green line) and direct numerical integration (red line). The value $F_1$ estimated in the previous paragraph allows us to read the corresponding photon fluxes $\hat{F}(F_1) \approx 1.5$ R for the Schechter function (green dot) and $\hat{F}(F_1) \approx 28.9$ R for the direct integration (red dot). The discrepancy between these two numbers shows the error introduced by using the Schechter function as compared to the direct numerical integration. We assess the direct numerical integration as a more reliable estimation because it takes into account a more accurate distribution of the point sources.

\subsection{Unresolved-point-sources anisotropy}
\begin{figure}[!htbp]
\begin{center}
\includegraphics[scale=0.55]{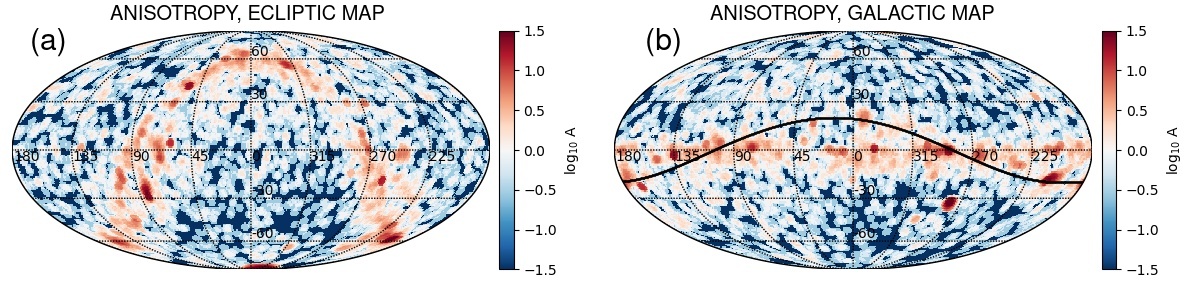}
\caption{Distribution of the anisotropy parameter $\log_{10}(A)$ in the sky in (a) ecliptic and (b) galactic coordinates. The color scale was chosen in such a way that white color corresponds to the local density of point sources equal to the all-sky-averaged density, redish colors correspond to the local density larger than the average and the blueish colors indicate regions with the local density smaller than the average. Black line in panel (b) shows the expected location of the Gould Belt, similarly to Figure \ref{fig:swan_map}(c) and (d).}
\label{fig:anis}
\end{center}
\end{figure}

The estimation provided in the previous subsection characterizes the background signal from unresolved point sources averaged over the entire sky. However, it is also interesting to investigate the anisotropy of the distribution of the point sources in the sky. To account for the anisotropy we introduce the parameter
\begin{equation}
A=\frac{4 \pi N_0}{ N \Omega_0}
\end{equation}
where $N=5615$ is the total number of point sources in the sky (a subset of the IUE observations used in our study) and $N_0$ is the number of point sources within the angular area $\Omega_0$ (we use the same value as above corresponding to the resolution limit $\rho_\mathrm{res} \approx 2.83$ deg). The parameter was constructed in such a way that for the local density of point sources $N_0/\Omega_0$ equal to the average density $N/(4\pi)$ the parameter $A$ is equal to 1, and regions where the local density is larger than the average density are characterized by $A>1$. In computations of the parameter $A$, both resolved and unresolved point sources are taken into account. Therefore, we compute the distribution of the anisotropy in the sky using all point sources, but the result of the computations is then interpreted as providing a reasonable approximation of the distribution of the unresolved objects. The distribution of $\log_{10}(A)$ is shown in Figure \ref{fig:anis}
in the ecliptic and galactic maps. Generally, regions of increased density of point sources are localized near the galactic-equatorial plane. The upper limit on the density of point sources was found to be $A_\mathrm{max} \approx 55$, which indicates a strong local anisotropy as compared with the average density of point sources.

In Figures \ref{fig:swan_map}(c) and (d) we show the position in the sky of the Gould Belt (black line), which is an apparent ring-like structure of alignment of bright-stars sequence that crosses the galactic equator at an angle $\sim\!\!20$ deg \citep{gould:79,guillout_etal:98}. Interestingly enough, it is difficult to discern the Gould Belt in the map shown in Figure \ref{fig:anis}(b), while such structure is clearly seen in Figures \ref{fig:swan_map}(c) and (d). The structure apparently extends over the range of longitudes from $\sim\!\!70$ to $\sim\!\!360$ deg, which suggests that the sequence of belonging stars covers a significant part of a great circle. The main difference between the maps shown in Figures \ref{fig:swan_map}(c),(d) and \ref{fig:anis}(b) is that in the former both the density of point sources and their brightness are taken into account, while the latter is based only on the density of point sources. This suggests some differences between the distribution of point sources and their brightness in the UV range of wavelengths 115-185 nm that can be interesting in the context of a debate on the nature of the Gould Belt structure.

The anisotropy of the distribution of point sources discussed in this section implies a question about the anisotropy of the unresolved-point-sources background. Since the results presented in this paper suggest a galactic-like pattern of arrangement for the background, we computed separately the average signal for the equator band (i.e., for the galactic latitudes between $\pm 20$ deg) that amounts to 49.7 R, the south ($<-20$ deg) -- 16.1 R, and the north ($>20$ deg) -- 16 R. In these computations an average anisotropy in all the regions of interest was found and the all-sky-averaged value of $28.9$ R was scaled using the average anisotropy value to obtain the average signal level for each region.

\begin{figure}[!htbp]
\begin{center}
\includegraphics[scale=0.65]{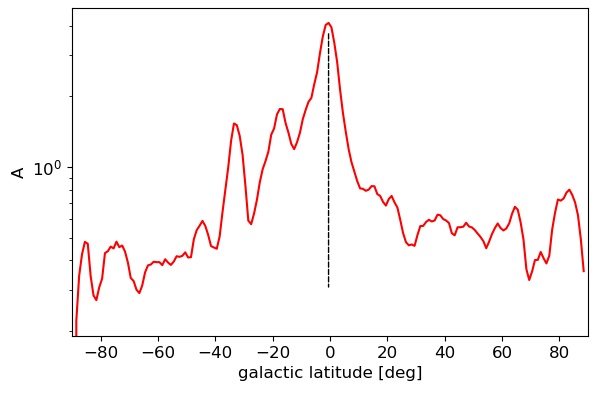}
\caption{Distribution of the galactic-longitude-averaged anisotropy. Vertical dashed bar in the center shows the location of the maximum of the distribution, which is shifted south by 0.5 deg.}
\label{fig:anis_long_avg}
\end{center}
\end{figure}

The sector-averaged anisotropy discussed in the previous paragraph gives very similar average values for the north and south part of the sky. However, the distribution of the anisotropy in the sky is not very symmetric with respect to the galactic equator if a more detailed picture is considered. Figure \ref{fig:anis_long_avg}
shows the distribution of the anisotropy averaged over the galactic longitude. The central peak is shifted slightly south by 0.5 deg but it is quite symmetric otherwise. The peak is relatively narrow as compared to the peak in the distribution of the photon flux shown in Figures \ref{fig:swan_map_long_avg}(a) and (c). Further from the galactic equator, in the range of latitudes from -40 to 0 deg, the anisotropy remains signicantly larger than in the corresponding sector in the north (0-40 deg), where it quickly drops below one. By contrast, closer to the poles, the contribution from the north is significantly larger than from the south.

\section{Discussion and Conclusions}
\label{sec:discussion}
We presented a general method to quantify a background radiation level from unresolved point sources in UV sky-survey maps. This universal framework can be applied to UV observations from any instrument, provided that the spectral sensitivity and angular resolution of the instrument are known. In the proposed method, a distribution of the number density of point sources is estimated from the IUE observations, but in general one can use any database providing a sufficient coverage of the region of interest in the sky. Then by proper integration of the distribution, a photon flux value is determined corresponding to the limit of the angular resolution of a given instrument. Based on this angular-resolution-limit photon-flux value, it is further possible to obtain an estimation of the average flux from faint point sources unresolved by the instrument.  This general method was applied to SOHO/SWAN observations, where the estimated all-sky-averaged photon flux from unresolved point sources amounts to $\langle F \rangle_\mathrm{UNRES} \approx 28.9$ R. The value results from the direct numerical integration of the distribution of point sources, that we asses as more accurate in comparison with the Schechter-function-based approach. Detailed discussion of the two approaches was presented in Section \ref{sec:avg_flux}. The estimated all-sky-averaged photon flux from unresolved point sources corresponds to approximately 3-8 \% of the heliospheric glow intensity as seen by the SOHO/SWAN instrument, depending on the solar cycle phase and direction in the sky.

Apart from the general all-sky-averaged estimation we also considered the question of anisotropy of the point-source distribution. The anisotropy shows strong variations in the sky, with extended regions of increased number density of point sources localized in the proximity of the galactic equator. When the anisotropy is taken into account and the sky is divided into three regions, the average unresolved-point-sources signal from the galactic-equator band is 49.7 R, from the southern sector -- 16.1 R, and from the northern region -- 16 R. Therefore, we found an increased emission from regions adjacent to the galactic equator, but no significant south-north asymmetry was identified. If we consider a more detailed picture, strong local anisotropies are observed in the sky, which is an interesting topic for further studies extending the coarse estimations that we discussed in this paper.

Based on the sky maps of point sources shown in Figure \ref{fig:swan_map}, we can compute the average intensity of signal for the SOHO/SWAN map [panel (a)] that gives $\langle F \rangle_\mathrm{SWAN} \approx 147.7$ R and for the IUE-based model map [panel (b)] that amounts to $\langle F \rangle_\mathrm{IUE} \approx 113.2$ R. The map in Figure \ref{fig:swan_map}(a) contains additionally the heliospheric-glow scatter of $\sim\!\!4.2$ R (estimated as a constant background in Figure \ref{fig:swan_map_long_avg}(b)) that needs to be subtracted to get the corrected value $\langle F \rangle_\mathrm{SWAN} \sim\!\!143.5$ R. The difference $\langle F \rangle_\mathrm{SWAN} - \langle F \rangle_\mathrm{IUE} \approx 30.2$ R represents the expected signal from point sources that are missing in the IUE database but they contribute to the sky survey provided by the SOHO/SWAN maps. The average unaccounted intensity is very close to our estimation of the unresolved-point-sources background. However, one must remember that in the computations of the average fluxes $\langle F \rangle_\mathrm{SWAN}$ and $\langle F \rangle_\mathrm{IUE}$, the PSF estimated in Section \ref{subsec:psf} is strongly involved, as the computations are based on the modeled map in Figure \ref{fig:swan_map}(b). By contrast, the averaged unresolved-point-sources flux $\langle F \rangle_\mathrm{UNRES}$ computations are based on simple multiplication by $\Omega_0$. Therefore, only a rough comparison between $\langle F \rangle_\mathrm{UNRES}$ and $\langle F \rangle_\mathrm{SWAN} - \langle F \rangle_\mathrm{IUE}$ can be done.

Since the unresolved-point-sources contribution depends on the spectral sensitivity of a given instrument, the estimations presented in this paper should be considered as specific for the SOHO/SWAN instrument. For other instruments the estimations can be different, but we assess the background is likely to amount to tens of Rayleighs, which is interesting, e.g., in the context of recent studies by \citet{katushkina_etal:17a}, where an additional emission of $\sim\!\!25$ R was identified as not accounted by models of the Lyman-$\alpha$ glow but obviously present in observations performed for an upwind field of view from the Voyager 1/UVS instrument. Our conjecture is that the unresolved-point-sources background may possibly contribute to the unaccounted emission seen by the UVS instrument.

We believe that the results presented in this paper contribute to better understanding of different components of the photon flux seen by the SOHO/SWAN instrument. The maximum of the helioglow intensity is located close to the local-interstellar-medium upwind direction and the minimum -- near the downwind direction. Both the upwind and downwind directions are placed in the proximity of the galactic equator, where the background signal from extra-heliospheric sources gives increased contribution. Therefore one can expect significant contaminations for the extrema of the helioglow intensity, which is important for comparison of the Lyman-$\alpha$ glow observations with models. Additionally, the upwind direction is useful for the analysis of ``pristine'' helioglow, where the solar photons are scattered on hydrogen atoms that have been least of all affected by the Sun. In this context, understanding of all components of the observed signal in the galactic-equatorial sector seems to be important. Our analysis suggests that systematic corrections for signal contributions from faint unresolved point sources are necessary for accurate interpretation of the Lyman-$\alpha$ glow. This study was performed in preparation for the GLOWS instrument onboard the planned IMAP mission \citep{mccomas_etal:18b} using currently available set of data for the Lyman-$\alpha$ glow.

\acknowledgments{The authors thank S. Grzedzielski for helpful discussions and M. Szpanko for providing guidance to the IUE data files. This work was supported by the Polish Ministry for Science and Higher Education. The authors also acknowledge partial support from NAWA grant PPI/APM/2018/1/00032/U/001 and from National Science Center, Poland, grant 2018-31-D-ST9-02852.}

\software{scikit-learn \citep{pedregosa_etal:11}, astropy \citep{astropy:13,astropy:18}}

\bibliographystyle{aasjournal}
\bibliography{iplbib}

\begin{thebibliography}{}
\expandafter\ifx\csname natexlab\endcsname\relax\def\natexlab#1{#1}\fi
\providecommand{\url}[1]{\href{#1}{#1}}
\providecommand{\dodoi}[1]{doi:~\href{http://doi.org/#1}{\nolinkurl{#1}}}
\providecommand{\doeprint}[1]{\href{http://ascl.net/#1}{\nolinkurl{http://ascl.net/#1}}}
\providecommand{\doarXiv}[1]{\href{https://arxiv.org/abs/#1}{\nolinkurl{https://arxiv.org/abs/#1}}}

\bibitem[{{Astropy Collaboration} {et~al.}(2013){Astropy Collaboration},
  {Robitaille}, {Tollerud}, {Greenfield}, {Droettboom}, {Bray}, {Aldcroft},
  {Davis}, {Ginsburg}, {Price-Whelan}, {Kerzendorf}, {Conley}, {Crighton},
  {Barbary}, {Muna}, {Ferguson}, {Grollier}, {Parikh}, {Nair}, {Unther},
  {Deil}, {Woillez}, {Conseil}, {Kramer}, {Turner}, {Singer}, {Fox}, {Weaver},
  {Zabalza}, {Edwards}, {Azalee Bostroem}, {Burke}, {Casey}, {Crawford},
  {Dencheva}, {Ely}, {Jenness}, {Labrie}, {Lim}, {Pierfederici}, {Pontzen},
  {Ptak}, {Refsdal}, {Servillat}, \& {Streicher}}]{astropy:13}
{Astropy Collaboration}, {Robitaille}, T.~P., {Tollerud}, E.~J., {et~al.} 2013,
  \aap, 558, A33, \dodoi{10.1051/0004-6361/201322068}

\bibitem[{{Astropy Collaboration} {et~al.}(2018){Astropy Collaboration},
  {Price-Whelan}, {Sip{\H{o}}cz}, {G{\"u}nther}, {Lim}, {Crawford}, {Conseil},
  {Shupe}, {Craig}, {Dencheva}, {Ginsburg}, {VanderPlas}, {Bradley},
  {P{\'e}rez-Su{\'a}rez}, {de Val-Borro}, {Aldcroft}, {Cruz}, {Robitaille},
  {Tollerud}, {Ardelean}, {Babej}, {Bach}, {Bachetti}, {Bakanov}, {Bamford},
  {Barentsen}, {Barmby}, {Baumbach}, {Berry}, {Biscani}, {Boquien}, {Bostroem},
  {Bouma}, {Brammer}, {Bray}, {Breytenbach}, {Buddelmeijer}, {Burke},
  {Calderone}, {Cano Rodr{\'\i}guez}, {Cara}, {Cardoso}, {Cheedella}, {Copin},
  {Corrales}, {Crichton}, {D'Avella}, {Deil}, {Depagne}, {Dietrich}, {Donath},
  {Droettboom}, {Earl}, {Erben}, {Fabbro}, {Ferreira}, {Finethy}, {Fox},
  {Garrison}, {Gibbons}, {Goldstein}, {Gommers}, {Greco}, {Greenfield},
  {Groener}, {Grollier}, {Hagen}, {Hirst}, {Homeier}, {Horton}, {Hosseinzadeh},
  {Hu}, {Hunkeler}, {Ivezi{\'c}}, {Jain}, {Jenness}, {Kanarek}, {Kendrew},
  {Kern}, {Kerzendorf}, {Khvalko}, {King}, {Kirkby}, {Kulkarni}, {Kumar},
  {Lee}, {Lenz}, {Littlefair}, {Ma}, {Macleod}, {Mastropietro}, {McCully},
  {Montagnac}, {Morris}, {Mueller}, {Mumford}, {Muna}, {Murphy}, {Nelson},
  {Nguyen}, {Ninan}, {N{\"o}the}, {Ogaz}, {Oh}, {Parejko}, {Parley}, {Pascual},
  {Patil}, {Patil}, {Plunkett}, {Prochaska}, {Rastogi}, {Reddy Janga},
  {Sabater}, {Sakurikar}, {Seifert}, {Sherbert}, {Sherwood-Taylor}, {Shih},
  {Sick}, {Silbiger}, {Singanamalla}, {Singer}, {Sladen}, {Sooley},
  {Sornarajah}, {Streicher}, {Teuben}, {Thomas}, {Tremblay}, {Turner},
  {Terr{\'o}n}, {van Kerkwijk}, {de la Vega}, {Watkins}, {Weaver}, {Whitmore},
  {Woillez}, {Zabalza}, \& {Astropy Contributors}}]{astropy:18}
{Astropy Collaboration}, {Price-Whelan}, A.~M., {Sip{\H{o}}cz}, B.~M., {et~al.}
  2018, \apj, 156, 123, \dodoi{10.3847/1538-3881/aabc4f}

\bibitem[{Bertaux {et~al.}(1995)Bertaux, Kyr{\"o}l{\"a}, Qu{\'e}merais,
  Pellinen, Lallement, Schmidt, Berth{\'e}, Dimarellis, Goutail, Taulemasse,
  Bernard, Leppelmeier, Summanen, Hannula, Huomo, Kehl{\"a}, Korpela,
  Lepp{\"a}l{\"a}, Str{\"o}mmer, Torsti, Viherkanto, Hochedez, Chretiennot, \&
  Holzer}]{bertaux_etal:95}
Bertaux, J.~L., Kyr{\"o}l{\"a}, E., Qu{\'e}merais, E., {et~al.} 1995, \solphys,
  162, 403

\bibitem[{{Bzowski} {et~al.}(2015){Bzowski}, {Swaczyna}, {Kubiak},
  {Sok\'{o}{\l}}, {Fuselier}, {Galli}, {Heirtzler}, {Kucharek}, {Leonard},
  {McComas}, {M{\"o}bius}, {Schwadron}, \& {Wurz}}]{bzowski_etal:15a}
{Bzowski}, M., {Swaczyna}, P., {Kubiak}, M., {et~al.} 2015, \apjs, 220, 28,
  \dodoi{10.1088/0067-0049/220/2/28}

\bibitem[{Edelstein {et~al.}(2006)Edelstein, Min, Han, Korpela, Nishikida,
  Welsh, Heiles, Adolfo, Bowen, Feuerstein, McKee, Lim, Ryu, Shinn, Nam, Park,
  Yuk, Jin, Seon, Lee, \& Sim}]{edelstein_etal:06}
Edelstein, J., Min, K.-W., Han, W., {et~al.} 2006, \apjl, 644, L153,
  \dodoi{10.1086/505208}

\bibitem[{{Gould}(1879)}]{gould:79}
{Gould}, B.~A. 1879, Resultados del Observatorio Nacional Argentino, 354

\bibitem[{{Guillout} {et~al.}(1998){Guillout}, {Sterzik}, {Schmitt}, {Motch},
  \& {Neuhaeuser}}]{guillout_etal:98}
{Guillout}, P., {Sterzik}, M.~F., {Schmitt}, J.~H.~M.~M., {Motch}, C., \&
  {Neuhaeuser}, R. 1998, \aap, 337, 113

\bibitem[{{Hunten} {et~al.}(1956){Hunten}, {Roach}, \&
  {Chamberlain}}]{hunten_etal:56}
{Hunten}, D.~M., {Roach}, F.~E., \& {Chamberlain}, J.~W. 1956, Journal of
  Atmospheric and Terrestrial Physics, 8, 345,
  \dodoi{10.1016/0021-9169(56)90111-8}

\bibitem[{{Johnston}(2011)}]{johnston:11}
{Johnston}, R. 2011, \aapr, 19, 41, \dodoi{10.1007/s00159-011-0041-9}

\bibitem[{Katushkina {et~al.}(2017)Katushkina, Qu{\'e}merais, Izmodenov,
  Lallement, \& Sandel}]{katushkina_etal:17a}
Katushkina, O., Qu{\'e}merais, E., Izmodenov, V.~V., Lallement, R., \& Sandel,
  B. 2017, \jgr, 122, \dodoi{10.1002/2017JA024205}

\bibitem[{MAST(2020)}]{mast:iue}
MAST. 2020, Mikulski Archive for Space Telescopes (MAST), IUE Data Products,
  \url{https://archive.stsci.edu/iue/file_formats.html}

\bibitem[{McComas {et~al.}(2018)McComas, Christian, Schwadron, Fox, Westlake,
  Allegrini, Baker, Biesecker, Bzowski, Clark, Cohen, Cohen, Dayeh, Decker,
  de~Nolfo, Desai, andH.A. Elliott, Fahr, Frisch, Funsten, Fuselier, Galli,
  Galvin, Giacalone, Gkioulidou, Guo, Horanyi, Isenberg, Janzen, Kistler,
  Korreck, Kubiak, Kucharek, Larsen, Leske, Lugaz, Luhmann, Matthaeus, Mitchel,
  Moebius, Ogasawara, Reisenfeld, Richardson, Russell, Sok{\'o}{\l}, Spence,
  Skoug, Sternovsky, Swaczyna, Szalay, Tokumaru, andP. Wurz, Zank, \&
  Zirnstein}]{mccomas_etal:18b}
McComas, D., Christian, E., Schwadron, N., {et~al.} 2018, \ssr, 214, 116,
  \dodoi{10.1007/s11214-018-0550-1}

\bibitem[{{Murthy}(2014)}]{murthy:14}
{Murthy}, J. 2014, \apjs, 213, 32, \dodoi{10.1088/0067-0049/213/2/32}

\bibitem[{Pedregosa {et~al.}(2011)Pedregosa, Varoquaux, Gramfort, Michel,
  Thirion, Grisel, Blondel, Prettenhofer, Weiss, Dubourg, Vanderplas, Passos,
  Cournapeau, Brucher, Perrot, \& Duchesnay}]{pedregosa_etal:11}
Pedregosa, F., Varoquaux, G., Gramfort, A., {et~al.} 2011, Journal of Machine
  Learning Research, 12, 2825

\bibitem[{{Quemerais} \& {Bertaux}(2002)}]{quemerais_bertaux:02}
{Quemerais}, E., \& {Bertaux}, J.~L. 2002, ISSI Scientific Reports Series, 2,
  203

\bibitem[{SCIKIT-LEARN(2020)}]{scikit-learn:svr}
SCIKIT-LEARN. 2020, Epsilon-Support Vector Regression,
  \url{https://scikit-learn.org/stable/modules/generated/sklearn.svm.SVR.html?highlight=svr#sklearn.svm.SVR}

\bibitem[{{Snow} {et~al.}(2013){Snow}, {Reberac}, {Qu{\'e}merais}, {Clarke},
  {McClintock}, \& {Woods}}]{snow_etal:13}
{Snow}, M., {Reberac}, A., {Qu{\'e}merais}, E., {et~al.} 2013, ISSI Scientific
  Report Series, Vol.~13, {A New Catalog of Ultraviolet Stellar Spectra for
  Calibration}, ed. E.~{Qu{\'e}merais}, M.~{Snow}, \& R.-M. {Bonnet}, 191--226,
  \dodoi{10.1007/978-1-4614-6384-9_7}

\bibitem[{SOHO/SWAN(2020)}]{soho:swan}
SOHO/SWAN. 2020, The SWAN Instrument onboard SOHO,
  \url{http://swan.projet.latmos.ipsl.fr}

\end{thebibliography}
%\bf{https://scikit-learn.org/stable/modules/generated/sklearn.svm.SVR.html}

\end{document}